\documentclass[twocolumn,amsmath,amssymb]{revtex4-1}
\usepackage{graphicx}
\usepackage{dcolumn}
\usepackage{bm}
\usepackage[usenames,dvipsnames]{xcolor}

\begin{document}
\preprint{ \color{NavyBlue} DRAFT v1.6 - \today}

\title{\Large \color{NavyBlue}  Bragg's reflection for walking droplets in 1D crystals}

\author{N. Vandewalle$^1$}
\author{B. Filoux$^1$}
\author{M. Hubert$^{1,2}$}
\address{$^1$ GRASP, Institute of Physics B5a, University of Li\`ege, B4000 Li\`ege, Belgium.}
\address{$^2$ \emph{(Current address)} PULS Group, Institut f\"ur Theoretische Physik, Friedrich-Alexander-Universit\"at Erlangen-N\"urnberg, 91052, Erlangen, Germany}

\maketitle


{\bf 
A walking droplet possesses fascinating properties due to its peculiar wave/particle interaction. The self-propelling motion of such a droplet is driven by the Faraday instability triggered around the droplet at each impact. We studied in this article how such a droplet behaves in an annular cavity where a periodic pattern is placed underneath the liquid-air interface, altering the Faraday instability. We show that, while the annulus ensures a circular motion of the droplet, the periodic pattern affects the global droplet motion. Similarly to electromagnetic waves in photonic crystals, the average droplet speed nearly vanishes when the pattern has a characteristic length close to half the Faraday wavelength. This effect opens ways to design guides, reflectors, lattices and metamaterials for such macroscopic particles.  
}


When a millimetric droplet is gently placed on an air-liquid interface vibrated with an amplitude $A$ and a frequency $f$ it is able to bounce without coalescing when the bath maximum dimensionless acceleration $\Gamma = {4\pi^2 A f^2 / g}$ is above a threshold $\Gamma_B$ \citep{Couder2005, Terwagne2013, Hubert2015}. An air layer separates the droplet from the vibrated surface preventing coalescence by lubrication \cite{Gilet2010}. Depending on $\Gamma$, various bouncing modes can be observed, from simple bounces to period doubling and much more complex bouncing dynamics \citep{Wind2013,Terwagne2013}. Once another threshold $\Gamma_F$ is reached, a parametric instability is triggered: the Faraday instability \cite{Kumar1996}. By approaching the Faraday instability from below, the droplet starts to bounce once every two periods, and triggers Faraday waves which possess a wavelength $\lambda_F$ being fixed by the forcing parameters \citep{Couder2005n,Protiere2006}. For $\Gamma_W < \Gamma < \Gamma_F$, due to the coupling between the droplet and the sum of waves emitted on the liquid surface by the successive previous impacts, bouncing droplets start to move horizontally along the liquid \cite{Couder2005n}. They are called walkers \citep{Protiere2006} and $\Gamma_W$ is named the walking threshold. The droplet-wave interaction leads to spectacular physical phenomena at a macroscopic scale: single walkers may exhibit tunneling effect \citep{Eddi2009}, diffraction and interference around apertures \citep{Couder2006}, revolution orbits \citep{Protiere2008}, splitting of energy levels \citep{Eddi2012} and quantification of macroscopic observables \citep{Perrard2014,Labousse2014}. Resonances \citep{Harris2013} and coherent wave packets \cite{Filoux2015} were also investigated above cavities. In all these reported phenomena, the major ingredient is the Faraday standing wave pattern which imposes a length scale $\lambda_F$ to the system. \\

In a recent study \cite{Filoux2017}, we demonstrated that the Faraday instability can be triggered locally by carving underwater cavities in order to guide the walking droplets along well defined paths. Wave-mediated transport phenomena can therefore be investigated. In particular, classical waves are deeply affected by the presence of a periodic media such that transport could vanish in some situations like in the Bragg's reflector case \cite{Bragg}. In this paper, we address the following fundamental question : what is the behavior of a walking droplet when perturbed by some underlying periodic pattern? \\

\begin{figure}[!h]
\vskip 0.1 cm
\includegraphics[width=8.5cm]{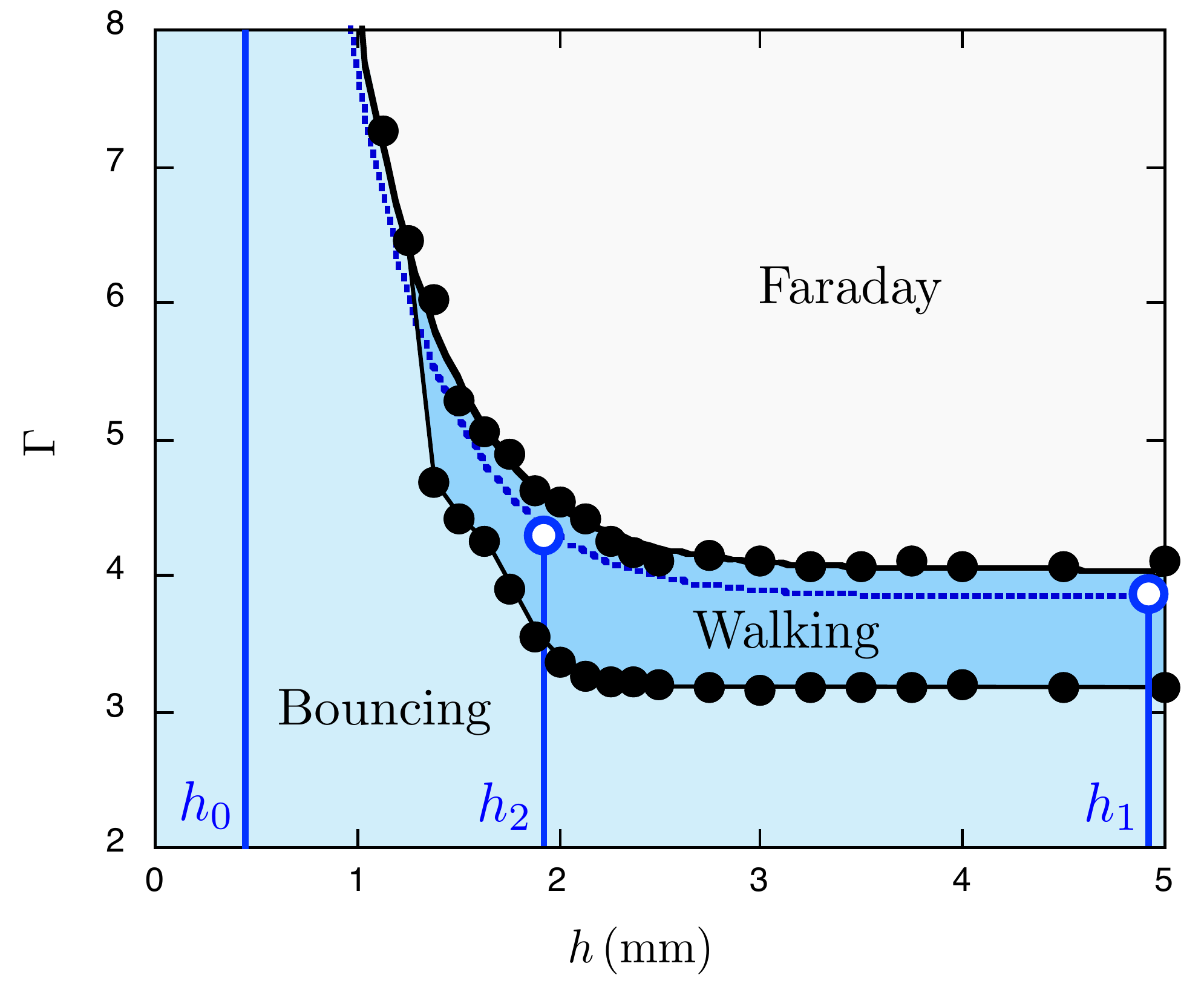}
\vskip -0.2 cm
\caption{{ \color{NavyBlue} \bf Phase diagram of dynamical regimes.} Dynamical regimes for the droplet depending on the reduced acceleration $\Gamma$ and oil depth $h$. Measured Faraday and Walking thresholds are indicated by filled circles. The dashed curve corresponds to Me=20. The vertical lines indicate the depths $h_0$, $h_1$ and $h_2$ fixed in our experiments. The empty circles correspond to the experimental conditions where the speeds $v_1$ and $v_2$ were measured.}
\label{pdf}
\end{figure}

The experimental conditions for the droplet and the bath are detailed in Methods. Identical droplets are created by an automatic generator described in Ref. \citep{Terwagne2013}. The liquid used in experiments is silicon oil. The container is shaken vertically by an electromagnetic system with a tunable amplitude $A$ and a fixed frequency $f=70 \, {\rm Hz}$  \cite{Filoux2017}. The dimensionless maximum acceleration $\Gamma$ is measured with an accelerometer and is adjusted to find the bouncing and walking regimes close to the Faraday instability. In an isotropic medium without any underwater carvings, the wavelength has been estimated to be about $\lambda_F^{2D} \approx 5.35 \, {\rm mm}$ in a large square container \cite{Filoux2015}. The characteristic of the Faraday instability are highly dependent of the liquid depth $h$ since the dispersion relation of gravity-capillary waves is given by 
\begin{equation}
\omega^2 = \left(gk+{ \sigma \over \rho} k^3 \right) \tanh (hk)
\end{equation}
which gives roughly the measured value of $\lambda_F = 2\pi / k$ when $hk \gg 1$, $h$ being the local depth of the fluid. Also, the occurence of the Faraday instability depends strongly of $h$ when $hk < 1$ given $\tanh (hk)$. Figure \ref{pdf} shows the Faraday threshold $\Gamma_F$, as well as the walking threshold $\Gamma_W$, for a droplet as a function of liquid depth $h$ in a large square cavity. Above $h=3$ mm, the behavior of both the droplet and the liquid surface are independent of $h$ while below $h=1$ mm, the bouncing regime vanishes and the Faraday regime only takes place on relatively important acceleration.   \\

\begin{figure}[!h]
\begin{center}
\vskip 0.1 cm
\includegraphics[width=8cm]{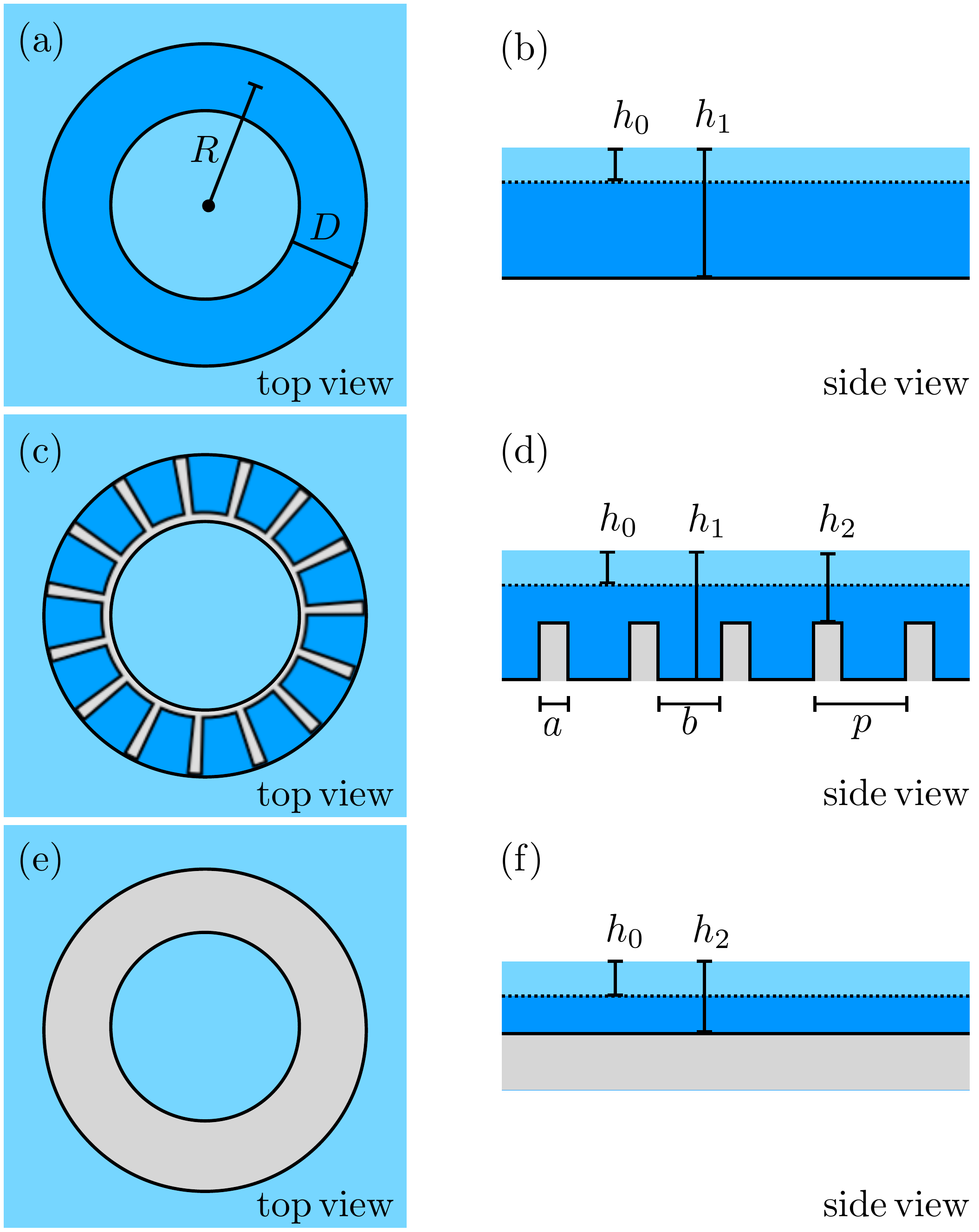}
\vskip -0.2 cm
\caption{{ \color{NavyBlue} \bf Sketch of the cavities.} (a) Top view of an annular cavity of width $D$ and radius $R$ is carved in the bottom of the oil container. (b) Side view. The oil level is adjusted to obtain a depth $h_1$ in the cavity and a thin layer $h_0$ elsewhere. By adjusting these parameters, a walking droplet tends to remain in the cavity and follow a circular motion. Additional cases were also considered. (c) A pattern of period $p=a+b$ is inserted in the annulus such that (d) zones of intermediate depth $h_2<h_1$ and width $a$ alternate with zones of depth $h_1$ and width $b$. (e,f) The case of uniform depth $h_2$ is also considered. }
\label{sketch}
\end{center}
\end{figure}

\section*{results}

Based on the results of Figure \ref{pdf}, we have proposed in recent works \cite{Filoux2015,Filoux2017} to study droplets in an annulus, as sketched in Figures \ref{sketch}(a,b). We decided to set the annulus mean radius to $R=13.75$ mm. The liquid depth in the center of the cavity is $h_1= 4.9 \, {\rm mm}$ and the width of the annular channel is $D = 7.5\, {\rm mm}$. This width has been demonstrated to be the optimal value for obtaining the best circular motion \cite{Filoux2017}. Elsewhere, the liquid depth $h_0$ is fixed to 0.4 mm, limiting the propagation of waves. Note that such shallow water areas are never explored by the droplet. Indeed, at this depth, a droplet may bounce but cannot walk as noticed in Figure \ref{pdf}. Observations show that typical trajectories of single walkers above the annular cavity of Figure 2(a,b) are circles. There is no central force behind the global circular trajectory. The fact that the droplet follows the annulus is that the Faraday waves adopt the symmetry of the cavity. In the azimutal direction, waves can be approximated to sinusoidal standing waves, as expected for a 1D system. In the transverse direction, i.e. in the radial direction, the waves present large amplitudes in the center of the cavity and evanescent waves are strongly damped outside the annulus. A relevant question here is to define and measure the Faraday wavelength in the present quasi 1D-system. Indeed, the presence of the annulus carved underneath the liquid surface affects the wavelength value. By measuring the quantified interdistances between droplet pairs moving along the same circular trajectory, we have demonstrated \cite{Filoux2015} that the natural wavelength triggered by droplets is $\lambda_F = 6.1 \pm 0.1 \, {\rm mm} $. Note that $\lambda_F \neq \lambda_F^{2D}$.\\

While the clockwise or anticlockwise droplet motion in the annulus depends on the initial conditions, the speed $v_1$ of a walker is fixed by the forcing parameters of the experiment as well as the geometry of the cavity \cite{Filoux2017}. In our experiments, the acceleration is always kept at $\Gamma = 0.95 \Gamma_F$. The Faraday waves are damped such that the liquid surface keeps the trace of about ${\rm Me} = \Gamma_F/(\Gamma_F-\Gamma) \approx 20$ previous impacts. For those experimental conditions, the walker speed is $v_1 \approx 6.6 \pm 0.3 \, {\rm mm/s}$. We will use this typical speed as a reference in the following. In order to affect the motion of the droplet similarly to Bragg's refractor, we place a periodic pattern inside the annulus to impose a different length scale, as sketched in Figures \ref{sketch}(c,d). It consists in a number $N$ of rectangular barriers of width $a$ separated by a distance $b$. The periodic length is therefore $2 \pi R/N = p=a+b$. On the barriers, the liquid depth is reduced down to $h_2=1.9$ mm. We measured the speed of droplets at Me=20 for this specific depth : $v_2 = 3.7 \pm 0.4$ mm/s, i.e. $v_2 \approx 0.56 v_1$. This specific depth corresponds to Figures \ref{sketch}(e,f). Assuming that each barrier is decreasing the speed of the droplet, we will consider periodic patterns for which the ratio $a/b$ is kept constant whatever the value of $N$. Therefore, we expect a time-average speed $\bar v$ given by
\begin{equation}
\bar v= {a v_2 + b v_1 \over a+b}
\label{eq:average}
\end{equation} whatever the number of barriers. Several periodic patterns have been created from $N=8$ to $N=44$ with a fixed ratio $a/b=0.25$. From the above equation, we expect $\bar v/v_1 \approx 0.91$. 

\begin{figure}[!h]
\begin{center}
\vskip 0.1 cm
\includegraphics[width=8cm]{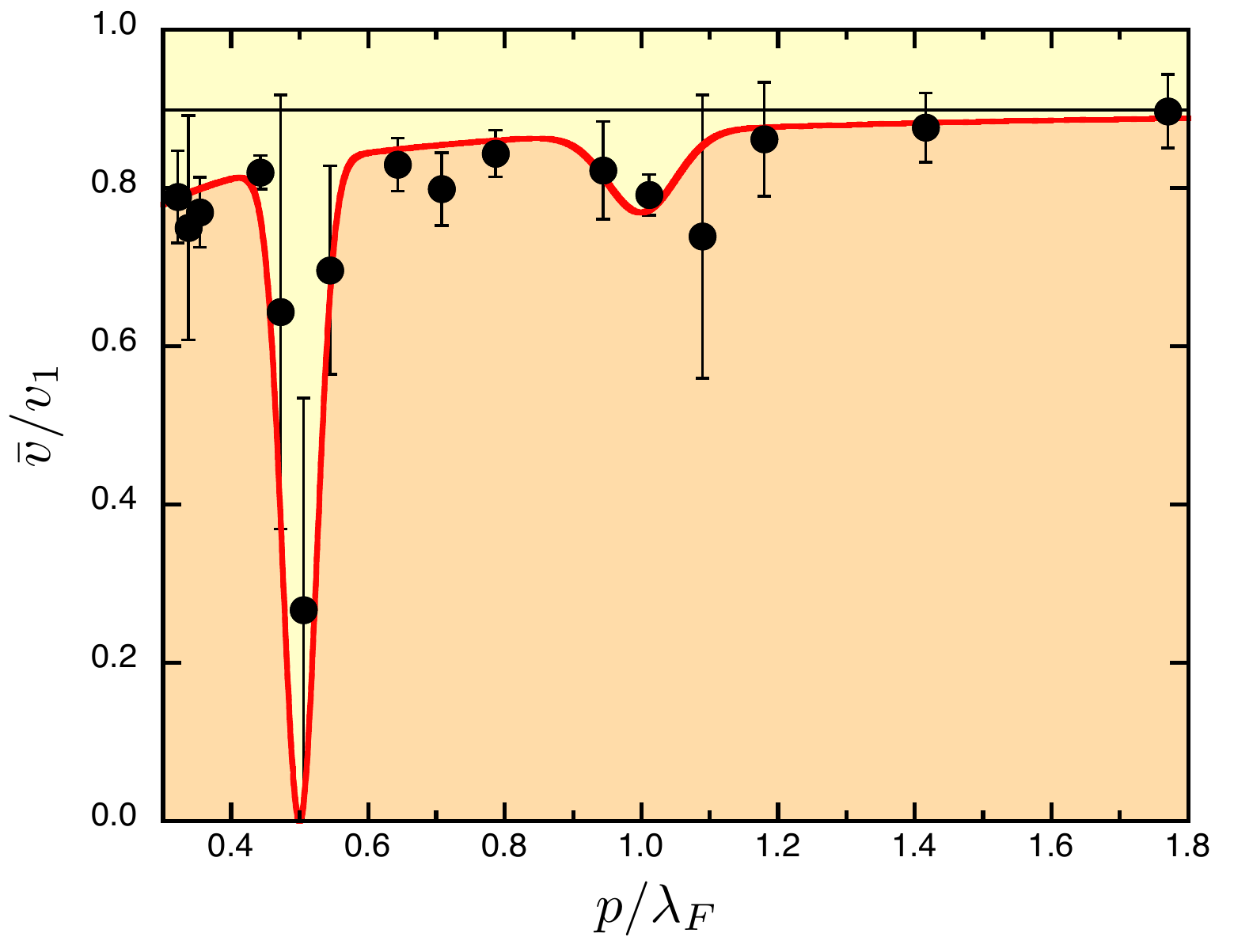}
\vskip -0.2 cm
\caption{{ \color{NavyBlue} \bf Speed of walkers.} Average normalized speed of walkers $\bar v/v_1$ in the annulus as a function of the period $p$ normalized by the Faraday wavelength $\lambda_F$. Error bars are indicated, measured over 20 different observations. The horizontal line corresponds to the expected average speed $\bar v/v_1$ from Eq.(\ref{eq:average}). The red curve is a guide for the eye, emphasizing the drop of average speed near the Bragg condition, i.e. Eq.(\ref{eq:bragg}).}
\label{speed}
\end{center}
\end{figure}

From about 400 independent experiments with constant forcing parameters, Figure \ref{speed} presents the average normalized speed of walkers $\bar v/v_1$ obtained by tracking the particles in the various cavities, as a function of the ratio $p/\lambda_F$. When $p$ is much larger than $\lambda_F$, the ratio is close to the expected value, given by Eq.(\ref{eq:average}), and indicated by the thin horizontal line. However, the average speed ratio is slightly decreasing with $p/\lambda_F$, due to the increase of the number of underneath barriers slowing down the droplet. We will obtain some evidence of that effect below. The striking feature of the plot is that one observes a sharp and significant drop of $\bar v/v_1$ meaning that the behavior of the droplet is deeply affected by the underlying pattern. It is worth noticing that such event takes place when approaching
\begin{equation}
p=a+b={\lambda_F \over 2},
\label{eq:bragg}
\end{equation}
which corresponds to the Bragg's condition \cite{Bragg}. The analogy between Bragg's reflection in optical lattices and our fluidic system is striking, even though major differences should be remarked. First, for classical light propagation in a 1D periodic layered medium, the wavelength is modified from one layer to the other one such that $p$ becomes the optical path slightly different from layer thickness. However, this is not the case herein since no wavelength variations have been observed. Second, the waves emitted by the walker are stationnary and does not propagate into the channel. Only their source (the walking drop) moves within the guide. Please note that higher order effects, at integer multiples of $\lambda_F$, have not been observed except that a small drop seems to take place around $p=\lambda_F$ with larger error bars. The red curve in Figure \ref{speed} is a guide for the eye emphasizing the main features of our results. \\

\begin{figure*}[t]
\begin{center}
\vskip 0.1 cm
\includegraphics[width=16cm]{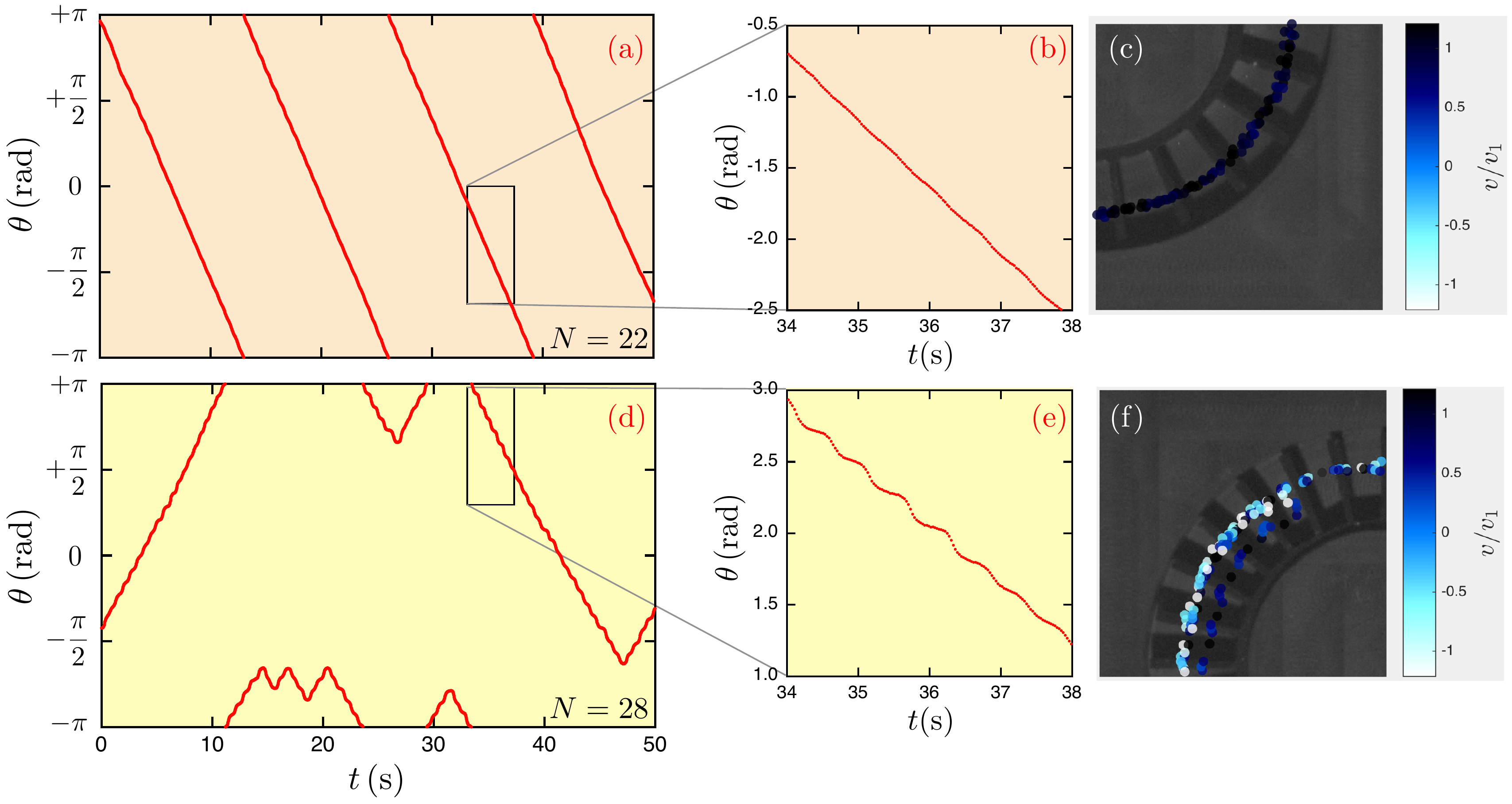} 
\vskip -0.2 cm
\caption{{ \color{NavyBlue} \bf Typical trajectories of walkers.} Azimutal trajectories $\theta(t)$ in the annulus for two different periodic patterns : (a) $N=22$ and (b) $N=28$, i.e. around the Bragg condition $p/\lambda_F=1/2$. In the former case, the droplet follows a circular motion at nearly constant speed, while in the latter case, a back-and-forth motion is seen to take place with randomness. For that condition,  the droplet speed is seen to oscillate when crossing periodically barriers, as shown in the enlarged parts of the plots (c) and (d). (e,f) Typical trajectories, corresponding to those conditions, with speed indication in color scale. This illustrates the fact that the speed decreases in between barriers. (f) For $N=28$, the speed is close to zero in between two successive barriers. }
\label{spatio}
\end{center}
\end{figure*}

To obtain a deeper understanding of this dynamics, figure \ref{spatio} presents the angular-time trajectories of the droplets in two cases: (a) $N=22$ (i.e. $p/\lambda_F = 0.64$) and (b) $N=28$ (i.e. $p/\lambda_F = 0.51$). The latter corresponds to the Bragg condition for which the average speed $\bar v$ vanishes. On both plots, the instantaneous speed are nearly constant. If one looks closely, one observes small undulations corresponding to the crossing of the successive barriers. Each time the droplet is passing a barrier, it decelerates and accelerates again. The speed oscillations are weak for $N=22$. However, for the case $N=28$, the droplet deceleration is so high that the droplet may stop and may turn back. Actually, the walking droplet makes random back-and-forth trajectories. Averaging this particular motion over long times leads to a vanishing average speed $\bar v$, while the instantaneous speed is nonzero. This behavior explains why large error bars are found in Figure \ref{speed} at the Bragg condition. A video is given in the supplementary materials in order to illustrate the particular droplet motion near obstacles. It seems that the droplet speed decreases at each obstacle crossing. The origin of the speed oscillations observed at the Bragg condition should be found in the Faraday wavefield around the walking droplet.\\ 

Let us now try to model this walking dynamics, and for this purpose let us look back at Fig.\ref{spatio}. Comparing Figs.\ref{spatio}(b) and \ref{spatio}(e), one sees that the short-time average speed of the walker changes depending on the value of $N$ (i.e. $p$). Indeed, in Fig.\ref{spatio}(b) the walker average angular speed within the channel is about $0.75$ rad/s while in Fig.\ref{spatio}(e) this speed does not even reach $0.5$ rad/s. Since the propulsive mechanism comes from the wave, one concludes that the presence of the submarine pattern has altered their generation. An explanation can be found in \cite{delon,Feng,Osipov}. In those studies, the authors focus on the appearance of Faraday waves with a submarine periodic pattern. It is shown that there exist wavelength of corrugated bottoms cancelling the Faraday instability, similarly to \emph{forbidden bands} in solid-state physics. The corresponding condition writes 
\begin{equation}
	\lambda = \lambda_F\frac{n}{2}, \quad n \in \mathbb{N},
\end{equation}
where $\lambda$ is the wavelength of the submarine obstacle. Please note that this condition has been obtained in a ``true'' channel. Our system only considers a channel carved underneath the surface: evanescent waves exist outside the channel boundaries. This observation explains why the walker still walks for ``Bragg's conditions'': because of boundary conditions in our experimental setup are smoother than in theoretical analysis, waves are still emitted but expected to be of smaller amplitude. The amplitude being smaller, the short term velocity of the droplet is also expected to be smaller, as observed in Figs. \ref{spatio}(b) and \ref{spatio}(e). \\

\section*{Discussion}

Let us consider the following numerical model for the walking dynamics. The numerical scheme is the one used in Ref.\cite{Borghesi2014,Labousse2016}. It consists in an event-driven algorithm where the walker motion is actuated in parallel of the wave dynamics. The walker alternates two phases: a parabolic flight above the liquid surface and a contact with the liquid surface. During the first phase, the droplet is only submitted to external forces while during the second phase, friction with the air layer leads to a loss of speed. In between, at each impact, a standing cylindrical wave is generated. Simultaneously, the walker gets a kick of momentum in the direction of the gradient of the total wavefield. A precise mathematical description of the algorithm can be found in \cite{Labousse2016}. Note that in those numerical simulations, the exact shape of the wavefield is assumed to be the superposition of waves with the shape $\zeta_0 J_0(2\pi\vert\vec{r}-\vec{r}_i\vert/\lambda)$ where $\zeta_0$ is the wave amplitude and $\vec{r}_i$ the center of the wave emitted. Therefore, the effect of the submarine obstacles are not taken into account and needs to be added by hand in the simulation. We use the ansatz theorized in \cite{Hubert2017}: we account for the submarine obstacle by using an external potential $U(r,\theta)$. The $r$ direction accounts for the confinement due to the annulus while the $\theta$ direction contains the periodic potential. As a consequence, we assume the following external potential
\begin{equation}	
	U(r,\theta) = U_0\cos\left(\frac{2\pi r\theta}{p}\right) + \frac{\omega^2}{2}(r-r_0)^2,
\end{equation}
where $U_0$ is the amplitude of the wavy external potential per mass unit. In this expression, the value of $\omega$ is chosen such as the width of the potential fits the width of the experimental channel, following the relation $\omega = d/\overline{v}$. In simulation, one uses $\pi$ $\mathrm{rad/s}$. Finally, we also known, as discussed before, that the wave amplitude decreases when approaching the Bragg's condition. As a consequence, the amplitude $\zeta_0$ is also expected to be changed.\\ 
\\

\begin{figure}[!h]
\begin{center}
\vskip 0.1 cm
\includegraphics[width=0.5\textwidth]{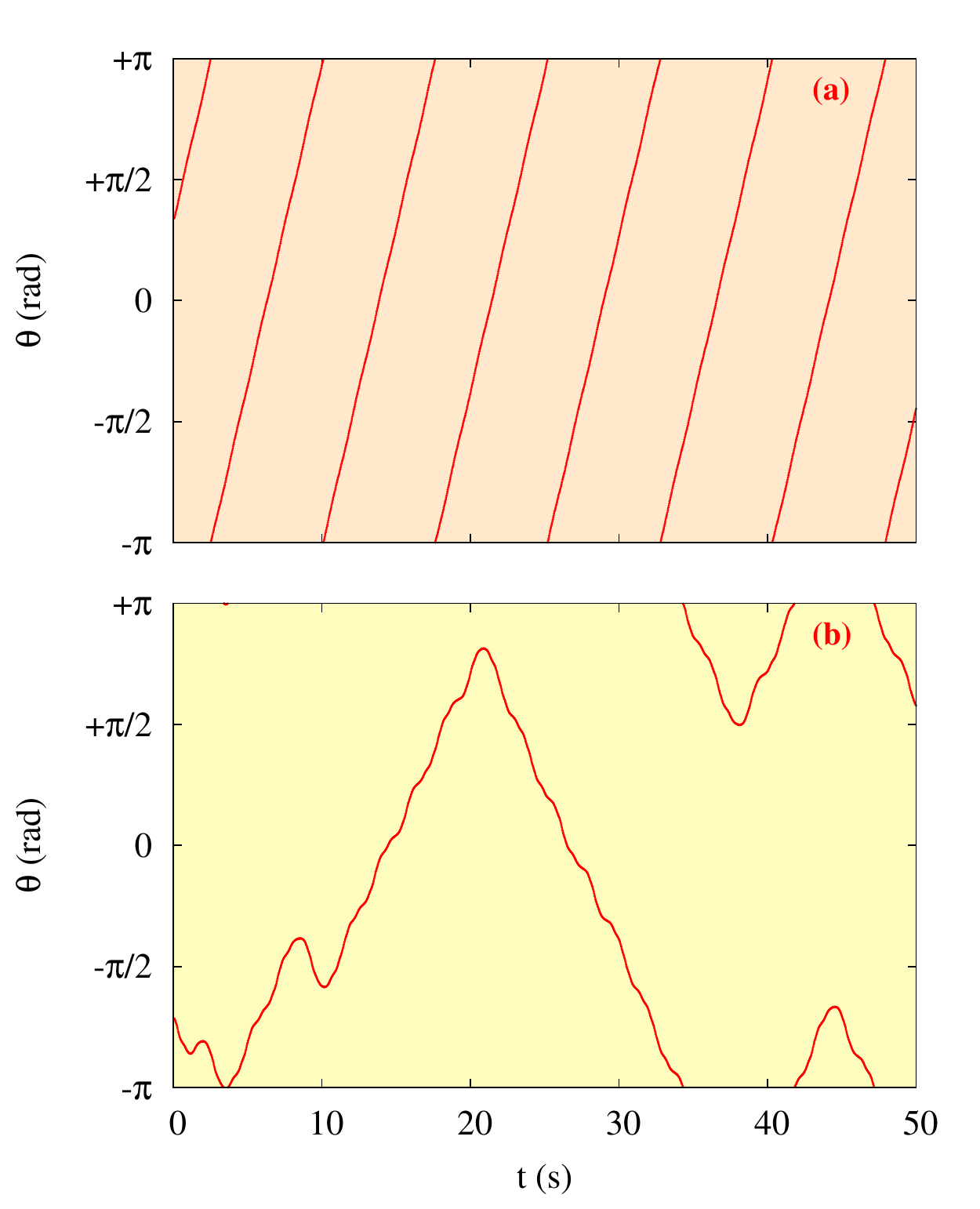} 
\vskip -0.2 cm
\caption{{ \color{NavyBlue} \bf Numerical simulations of the walkers dynamics within corrugated medium.} (a) The amplitude of each individual wave source is set to $\zeta_0$. (b) the amplitude of each individual wave source is $0.66 \zeta_0$. For each simulation, the amplitude of the wavy external potential is $U_0 = 15$ $\mu$J/kg, which is comparable to the energy along the $y$ direction. The periodicity of the underwater carvings is identical and corresponds to $p=\lambda_F/2$.}
\label{numSpatio}
\end{center}
\end{figure}

Figure 5 illustrates the dynamics of the walker in numerical simulation for two different amplitude of emitted waves, namely $\zeta_0 = 5$ $\mu\mathrm{m}$ (a) and $0.66 \zeta_0$ (b) and identical periodicity $p = \lambda_F/2$. The wavy external potential is $U_0 = 15$ $\mu$J/kg, which is comparable to the energy along the $r$ direction. The equilibrium radius $r_0$ has been chosen such that it matches the experimental one. Note that the memory has been kept constant in each simulation with $\mathrm{Me} = 20$. One can see that the dynamics observed mimics the experimental one, validating our hypothesis: a change of the wavefield amplitude indeed allows the walker to be reflected by the external potential, similarly to the studies of Eddi et al \cite{Eddi2009} and later explained theoretically by Hubert et al \cite{Hubert2017}. As the amplitude of the wavefield gets smaller at the Bragg's condition, the active mechanism propelling the droplet gets weaker. In this condition, when the droplet comes forward a shallow area region, it acts as a repulsive potential which might repeal the walker. This simple numerical experiment allows for three conclusions. First, the dynamics shows in Fig.3 only comes from the hydrodynamic and can be explained through the appearance of the Faraday instability. Secondly, this is the walker weaker active mechanism (and therefore lower speed) that explains the reflection above submarine obstacles. Finally, one has to understand that contrarily to ``true'' Bragg reflection, the walker dynamics changes because of a change of wave amplitude instead of a change of wavelength.\\

In summary, we evidenced from experimental measurements that a Bragg condition exists for the transport of droplets above periodic patterns. This effect is seen to originate from the pinning of the Faraday standing wave on the underneath pattern. This Bragg-like remarkable phenomenon can be exploited for creating resonators, metamaterials and other systems for guiding walking droplets.


\section*{Methods} The experimental conditions for the droplet and bath are the following. Identical tiny droplets are created by an automatic generator as fully described in \citep{Terwagne2013}. The resulting diameter of each droplet is close to 800 $\rm \mu m$. The liquid is silicon oil with a kinematic viscosity of 20 cSt, density $\rho=949 \, {\rm kg \, m^{-3}}$ and surface tension $\sigma = 20.6 \, {\rm mN \, m^{-1}}$. The container is shaken vertically by an electromagnetic system with a tunable amplitude $A$ and a fixed frequency $f=70 \, {\rm Hz}$. An accelerometer is fixed on the vibrating plate and delivers a tension proportional to the acceleration. The dimensionless maximum acceleration $\Gamma$ is tuned for finding the bouncing and walking regimes close to the Faraday instability. The acceleration is always kept at $\Gamma = 0.95 \Gamma_F$ in our experiments. This corresponds to a so-called ``low memory regime" since Faraday waves are damped such that the liquid surface  keeps the trace of a few previous impacts. The wavelength has been estimated to about $\lambda_F \approx 6 \, {\rm mm}$.\\

\vskip 0.2 cm
{\bf Acknowledgments} -- This work was financially supported by the Actions de Recherches Concert\'ees (ARC) of the Belgium Wallonia- Brussels Federation under Contract No. 12-17/02. We thank also S.Dorbolo for fruitful discussions.

\vskip 0.2 cm
{\bf Supplementary Information}

A movie is given : a droplet moving in the annulus were a periodic pattern of $N=28$ barriers is placed, fulfilling the Bragg condition. The droplet has a random back-and-forth motion.

\vskip 0.2 cm
{\bf Authors Contributions Statement} -- BF collected and analyzed experimental data. Physical interpretations were provided by BF, MH and NV. Simulations were performed by MH. This manuscript was written by MH and NV. 

\vskip 0.2 cm
{\bf Competing Interests} -- The authors declare that they have no competing financial interests.


\end{document}